# Using Descriptive Analytics for the Improvement of National University Entrance Exam: A case study in the Context of Kankor in Afghanistan

**Abdul Rahman, Sherzad**

**Lecturer at Computer Science faculty of Herat University, Afghanistan**

**Ph.D. Student at Technische Universität Berlin, Germany**

absherzad@gmail.com | absherzad@mailbox.tu-berlin.de

## Abstract

In Afghanistan, High school graduates, in order to continue higher education, need to pass the National University Entrance Exam (Kankor, from the French word Concours; English: "Contest"). Kankor is very important and requires further research and studies as it is used as the only means and tools to identify the participant's competence and skills which leads to a qualified higher education and manpower, employment opportunities, allocation of human capital in appropriate fields of study, and leading to a more specialized labor force in the long run.

Having a better picture of Kankor in Afghanistan is essential before employing educational data mining techniques and other social and pedagogical approaches to support both the students and the educational institutions. Hence, the author carried out a research on Kankor, in which he assessed and analyzed Kankor from the following angles: candidates' performance, the performance of public and private high schools, candidates' choice of field of study, the contribution of high school performance and grades, etc. This research paper provides positive contribution to the Ministry of Higher Education (MoHE), particularly, the Kankor committee, and other educational institutions – introducing the importance of data as a valuable asset and a proper structure of recording and producing data effective for research studies, paving the way for in-depth descriptive and predictive analytics which improves the situation within the context of the existing structural models in Afghanistan.

The author's research target is to study Kankor using descriptive analytics supported by Kankor Results Data (KRD) of 1.5 million records from 2003-2015, Detailed Kankor Results Data which is a subset of KRD from 2004 – 2006 of 120,000 records, High school performance and grades data from 2011-2013 of 6,000 records, data collected through conducting more than 2500 questionnaires among academicians as well as the author's personal observations, findings, and analyses as a student and faculty member at the University in the field of computer science.

# Introduction

A plethora of researches conducted by credible international entities in Afghanistan after the collapse of the Taliban regime reveal that investment in primary, secondary and tertiary education is vital to the sustainability of an inclusive development. Similar researches carried out in post-conflict nations and third-world countries also show similar results. While the past decade's quantitative and structural developments are considerable, Afghanistan continues to be ranked the lowest in South Asia. In Afghanistan, most of the policymakers share common characteristics, in the sense that everyone wants to be result-oriented without thinking about the order of effects of their policies or decisions. A bright future for Afghanistan is only possible when its citizens and public servants know that major problems were only caused by a collection of smaller factors and only a reverse similar behavior is required to resolve the given situation.

The statistics, available literature, and data indicate that more than 70% of participants severely lack a proper understanding of the structure, methods and overall procedures for the Kankor test. More than 80% believe unfamiliarity with Kankor test and lack of preparation for Kankor are two of the main reasons participants fail in the Kankor. Besides, high school education lacks the fundamental structure of a counseling system to orient students with regards to students' career choices and relevant fields of study. Therefore, more than 90% of Kankor candidates participated piloted Kankor test for the purpose of self-assessment, and to understand Kankor process and test.

Kankor, as a standardized test is found not to be the appropriate medium to identify the level and extent of participants' knowledge. While 50% of the questions focus on mathematics and science, the other 50% is allocated to the rest of areas of study including languages, geography, history, theology, etc. For example, some participants who are interested to follow a degree in Fine Arts fail to pass Kankor, because half of the Kankor questions are mathematics and science questions with higher scoring value. Some also get admitted into other Majors since admission to higher education is highly dependent on the number of available slots or the general absorption capacity of the higher education institutions. These structural factors pose a serious threat to appropriate division of labor if Afghanistan is to go the right path in the long run. On the other hand, structural deformities in Afghanistan's education system highly depend on political policy reforms, which might not be practical unless the ground is paved to create that willingness. Thus, the idea is to propose a solution, which capitalizes on improving the situation within the context of the existing structural models. This position paper will support the Kankor committee of MoHE, and other relevant educational institutions in the following domains:

- This paper discusses the value of data and accordingly proposes a proper structure to store and produce detailed data in order to enable policymakers to produce accurate and comprehensive insights and interesting patterns supported by facts and figures,

- This paper will ensure a better ground for further future educational data mining applications to improve the settings effectively and efficiently,
- The outcome of this research leads educational institutions to consider the order of effects in their policies and decisions.

In brief, the outcome of this research could act as a valuable framework and guideline for MoHE to enrich and support the Kankor settings effectively.

## Kankor

The MoE and MoHE are the two institutions directly responsible for addressing the educational needs in Afghanistan. In order to enter higher education, high school graduates need to pass Kankor.

Kankor is held every year, mainly in the capital and large provinces, usually between December and the end of February. Since 2006 the Kankor examination process was computerized – the questions bank and paper setting, Optical Mark Recognition (OMR) form correction and assess the answer sheets. Prior to 2006 usually 6 months were required in order to finish Kankor – from registering for the exam until announcing the final results (Sakhi & Nabizadah, 2015, p. 9). The Kankor exam comprises 160 questions about subjects taught at high school mainly in grades 10 to 12 categorized into the following categories: Mathematics, Natural Science, Social Science, and Languages. Kankor questions are provided both in Pashtu and Dari, in Pashtu and Dari speaking regions respectively (MoHE, 2011, pp. 45–56). All the questions are multiple choice, and generally, the participants have two to three hours to answer and choose the proper options. Usually, correct answers are worth one to three points. The maximum number of points are between 320 and 370, but more important is the minimum number of points needed to qualify for a university slot. Applicants can choose five favorite fields of study, and each field of study requires a certain score. In general, admission into popular fields of study like medicine, engineering, computer science, economics, and political sciences and into institutes in the capital and main cities requires higher scores. However, if a student does not score high enough for any of his/her chosen fields of study, he/she is dropped altogether and is not assigned any field of study, a result called result-less (benatedja). Moreover, a candidate can perform well and score high in Kankor, yet be unlucky and still not get into his/her most desired field of study. Because admission also depends on the number of available seats at the target higher education institutes (Ali, 2015).

In 2003 25,424 students participated in Kankor. 14,283 of them were accepted into public universities. In 2014 and 2015, there were 261,064 and 181,836 Kankor participants. The number of participants accepted into public universities was 66,024 and 56,729 respectively. The increase in the number of high school graduates is explicitly disproportional to the rate of acceptance in the Kankor, almost 1/5. Each year this gap inevitably grows larger, leaving more candidates frustrated. It is very challenging to get admitted into the desired higher education institutions when more than 250,000 high school graduates are applying for about 55,000 slots. Additionally, since 2011, a Kankor volunteer only has two chances of attending Kankor in a lifetime while before they had three chances.

In order for MoHE to address this challenge, policymakers expanded the role of semi-higher education institutions where eligible Kankor participants are introduced to Teacher Training, Islamic and Technical and Vocational institutions. Despite this, the problem continued and since 2013, MoHE decided to introduce a large portion of the eligible participants to private universities; with a discounted fee – if it is affordable for them. In order for Kankor participants to cope with this challenge, those who can afford it attend supportive courses and invest in private Kankor preparation courses and popular private schools. This is mainly due to continuing insufficient instruction in high schools and especially the lack of qualified teachers (particularly in sciences). These private supportive courses are usually more efficient and are run mostly by qualified teachers and university graduates.

## Research Methodology

This paper uses a mixture of both quantitative and qualitative research methods. No such research and data collection methodology have been used in the past to evaluate the technological context in Afghanistan, particularly in the academic field, thus making the research unique and an added-value in the body of research in this area.

1. Kankor Results Data (KRD) from 2003 – 2015 containing 1.5 million records of Kankor participants' personal information and their performance in the Kankor as well as the result of their performance (admission to the public university, semi-higher education, private university, or fail). The KRD is used for the purpose of descriptive analytics.
2. Detailed Kankor Results Data (DKRD) from 2004 – 2006 is a subset of KRD with around 120,000 records of Kankor participants. The DKRD contains additional data e.g. Kankor candidates' scores for Languages, Mathematics, Natural and Social Sciences as well as candidates' successful acceptance into desired fields of study. The DKRD is used to find out how much Mathematics and Science scores play a positive role in a candidate's acceptance. Also, to assess whether a higher score always plays a role in candidate's successful admission to his/her first choice of field of study.
3. School Performance Data (SPD) from 2011 – 2013 containing around 6,000 records of high school graduates' information as well as subject-wise grade score for 10th, 11th, and 12th grades. The SPD, after matching with KRD, is used for descriptive and predictive analytic purposes to find out if the high school performance plays a positive role in a candidate's successful admission to higher education.
4. A research conducted among 1,309 public and private university and high school students through questionnaires to know 1-where they take preparation for Kankor and how much it is helpful for passing Kankor, 2-before Kankor, how much are/were they familiar with Kankor test and procedures, 3-what are the main reasons students failing in the Kankor, 4-how much Kankor piloted practice tests will be useful. A similar research conducted among 632 respondents in public and private universities and 603 respondents in public and private schools to know: 1- whether Kankor is a proper means to identify the candidate's skills and proper field of study, 2-how much knowledge they have/had about the Kankor test, procedures and policies before taking it, 3- how much they knew about the field that they are supposed to study and on what basis they choose field of study in Kankor, 4-how much counseling and advising on career choice positively impacts proper selection of field of study and what methods they suggest.
5. This research also relies on the personal observations, findings, and analysis of the researcher as a student of computer science and later as an active and experienced lecturer and academic member at the University in the field of computer science.

## Data Preparation and Preprocessing

Data collection and data pre-processing to detect and correct/remove corrupt or inaccurate data from a dataset are the very challenging parts given that there are no established means of collecting data and the practice of regular data collection are lacking. The author identified and corrected inconsistent data, detected and addressed outliers and smoothed out noisy data, filled in missing value, and then combined data from multiple sources into a coherent dataset (Osborne, 2012). It is worth it, as it leads to transformation of data into an organized structure and generation of accurate, valuable and detailed insights.

### Kankor Results Data from 2003 - 2015

First, a general structure with the following attributes (ID, Kankor Year, Kankor ID, First Name, Family Name, Father's Name, Grand Father's Name, Gender, High School Name, High School Graduation Year, Province, Kankor Score, Candidate's Result) was created.

Then, the data inconsistencies were identified and properly addressed: For example, province names appear in different formats such as (Balkh, Mazar, and Mazar-e-Sharif) using different naming conventions for the same province. Also, the gender column was identified with different values (i.e. Male, M, 0, and other different equivalent Persian identifiers) all as male gender identifier. Furthermore, invalid data and typo were detected and corrected: For example, the Kankor is for the year 2003, but there were participants with school graduation year of 2007 and 2011 which is impossible.

Afterward, the missing values for the attributes useful for statistics i.e. Gender and Province of the candidates' school were identified and filled with correct values. To address missing value for gender and school location, the whole datasets first on "First Name" and then on "School Name" were sorted out respectively. For the majority of the candidates, gender and school location were already identified and based on the already available information, the missing values were filled accordingly.

After that, in order to provide a better insight into the data, the "Candidate's Result" attribute was divided into the following four attributes (Field of Study, Educational Institution, Type of Educational Institution, and Location of Educational Institution). For example, "Software Engineering at Computer Science faculty of Kabul University only for male", broken down into "Computer Science" as Field of Study, "Kabul University" as Educational Institution, "University" as Type of Educational Institution, and "Kabul" as Location of Educational Institution. It is worth mentioning that typo and different naming conventions were present and corrected during this division process: For example, computer science, computer education, computer engineering, computer and networks, computer and IT, software engineering, and other conventions were replaced to "Computer Science". It is worth mentioning that the location and type of some of the educational institutions could not be identified even with advanced filtering techniques. On those special cases, the author's familiarity with the education structure and Google were useful. Thus, all the conventions were unified.

Next, attributes useful for generation of facts and figures were transformed into their equivalent English terminology. Even the province attribute for candidate's school and the location attribute for educational institutions were also categorized into Region sections.

Finally, the private portion of the data including First Name, Family Name, Father's and Grand Father's Name, and other confidential and personal information was excluded to ensure the candidate privacy.

### School Performance Data 2011 - 2013

Around 6,000 records of school performance of high school students from 2011 to 2013 were collected. The students' personal information and the scores for 10th, 11th, and 12th grades data were stored separately in different entities and based on the primary key, data was merged together. After this process, 1,500 duplicated data were found which was addressed through this rectification process. Then this dataset was compared with KRD based on the following attributes (First Name, Family Name, High School Name, Province, and High School Graduation Year) to find out who had participated in the Kankor exam to evaluate the impact of their high school grades on their Kankor results. Around 1,800 records of the students who had participated in the Kankor were found. Finally, after validating the data, the data from school performance were merged together with KRD and a new dataset with 1,800 records was formed.

# Results of Descriptive Analytics

Data is a valuable asset to any organization. Graham Williams (2011, p. 57) mentions in his book that "Data is the starting point for all data mining – without it there is nothing to mine". Also, there is a quote from W. Edwards Deming "Without data you are just another person with an opinion". Statistics as a scientific discipline provides methods to help organizations make sense of data and gain significant insights through exploring the data. These insights can deliver new discoveries that can offer benefits to policymakers, and in a data mining projects. Through such insights and discoveries, the organizations will increase their knowledge and understanding (Peck & Devore, 2011; Williams, 2011). Correct and clean data directs to accurate and transparent insights, reports supported by facts and figures. More detailed data gives organizations the opportunity to try to figure out much more interesting patterns and leads to more specific and significant analytics and insights. In this section, the author explains a few examples from the descriptive and exploration analytics perspectives. This will support the organizations and policymakers to understand the 'lay of the land', and in turn, will drive the choice of the most appropriate and significantly applicable tools for preparing and transforming the data for future educational data mining applications to improve the settings effectively and efficiently.

## Accurate Facts and Figures

With proper tags and identifiers of the following attributes: gender, field of study, educational institution, type and location of educational institutions, etc. accurate facts and figures could be easily generated: For example, to precisely produce candidates' admission based on type of institutions i.e. public university, private university, Technical and Vocational institutions, Teacher Training, and Islamic Studies (see Figure I). It is worth mentioning that since 2013 admission to private universities substitute rates of failure.

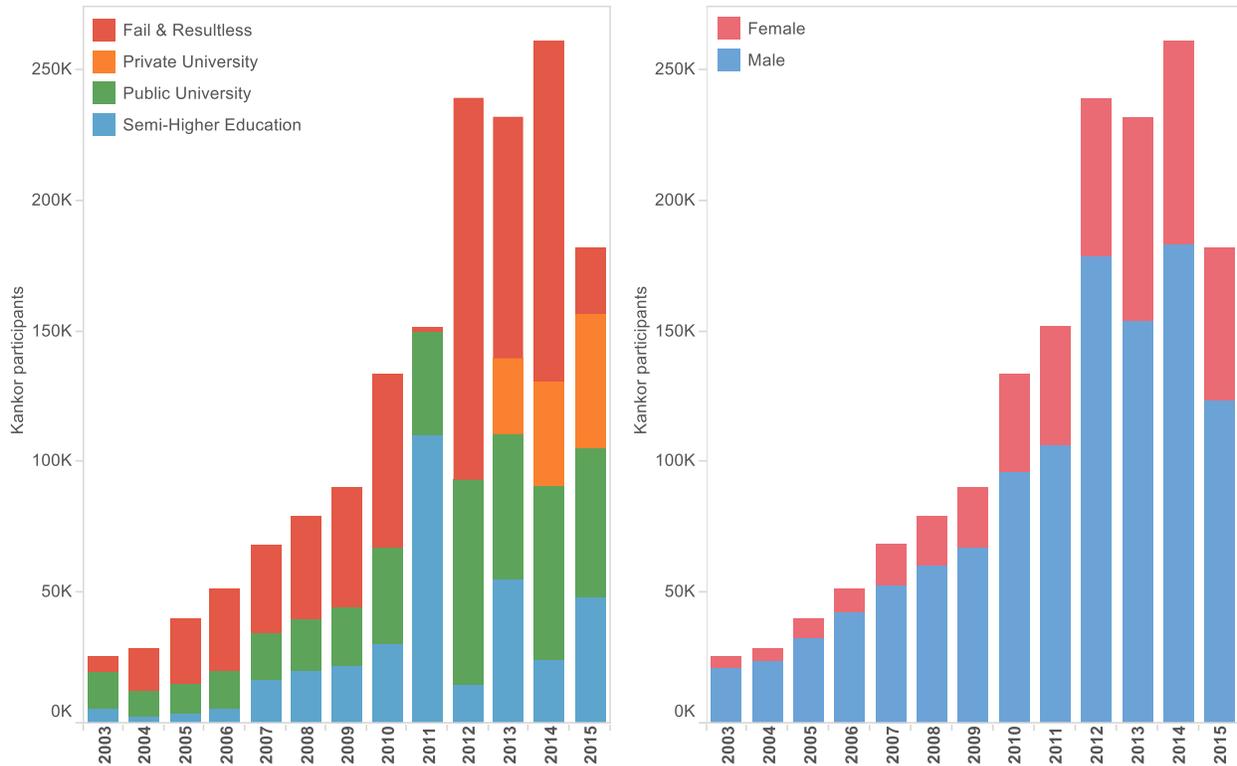

*Figure 1: Kankor participants' trends Result-wise (left chart), and Gender-wise (right chart) since 2003*

The conditions could be grouped to produce further specific and detailed insights: For instance, to generate the top-n higher educational institutions and fields of study with high enrollment trend location-wise, gender-wise and using other criteria for evaluation and assessment purposes. The author's experimentation and analyses indicate Teacher Training College and Faculty of Education (Pedagogy) are with high enrollment rate and trend across Afghanistan (see Figure 2). Despite this still primary, secondary and high schools suffer from the lack of qualified teachers resulting in a poor education. Several factors could cause this: low quality in the higher education institutions, the recruitment process for hiring of school teachers is not transparent, or it could be another reason which might require further investigation. Likewise, although Faculty of Agriculture and Technical Institute of Agriculture are ranked the second top (see Figure 2), Afghanistan still suffers from not having experts in the field. Finally, statistics also show top enrollment in technical and vocational institutions (see Figure 2). Unfortunately, unemployment and lack of professionals are still noticed and have increased across the country. These provide an opportunity to policymakers to investigate the causes. Additionally, one can find the fields of study with more or less female enrollment based on the Kankor year and the province. This will provide the opportunity for policymakers to increase female enrollment in those fields of study to boost women's role in the society in the future.

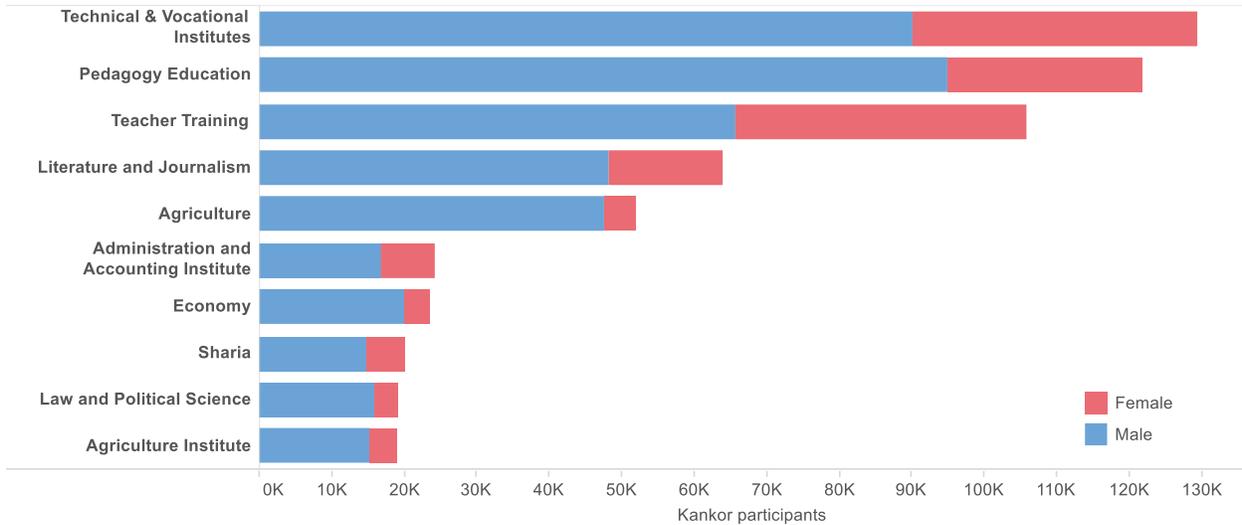

*Figure 2: The top 10 fields of study in Afghanistan higher education institutions with high enrollment trend since 2003*

## Candidates Performance Assessment

Statistics and analyses illustrate that most candidates score below 200 in Kankor. Moreover, there are many candidates with good performance and high scores who are introduced to private higher education institutions rather than public universities and vice versa (see Figure 3). It can be concluded that admission also depends on how many slots are available at the different educational institutions, thus, it is "first come first served". A candidate can perform well and get a high score yet be unlucky and still not get into his/her most desired field of study. Therefore, the candidate's unknowledgeable choice of field of study might lead to unsuccessful admission.

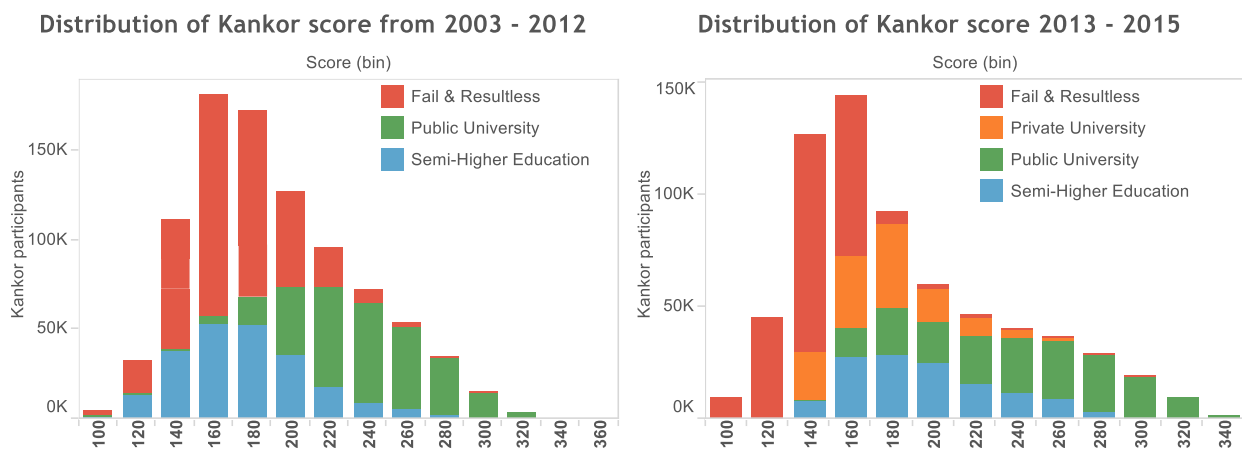

*Figure 3: Distribution of participants' Kankor score and their result*

## Public and Private High School Performance Assessment

The author chose three high schools with high reputation from Herat province, two public high schools, and one private high school. Analyses revealed candidates from private high school perform better than candidates from public high schools in the Kankor. It is worth mentioning the selected private high school has different strategies as it accepts (only) the top and skilled students after they successfully pass different tests and competitions. But all private high schools do not perform well in Kankor. This was confirmed using data from SPD and KRD. The author proposes that if proper tag is introduced for high schools and other information i.e. public or private, urban or rural, etc. then, not only ratio of successful Kankor participants based on types of high schools could be generated, but also the question of whether private high schools prepare the Kankor participants better than public high schools could be answered more precisely. Then the policymakers can provide proper solutions to improve the settings.

## Candidates Choice of Field of Study Assessment

The candidates could opt for ten fields of study they favor, and for one location for each field of study. They also had three chances to participate in Kankor in lifetime until 2011. Since then, they can opt for only five fields of study and have only two chances in the lifetime. The author evaluated candidates' acceptance based on their choice of field of study in the Kankor. It can be concluded that some candidates with good performance and score did not get admitted into their first favorite choice of field of study. On the other hand, some candidates with lower performance and score did (see Figure 4). Thus, it can be said that proper choice of field of study is very critical.

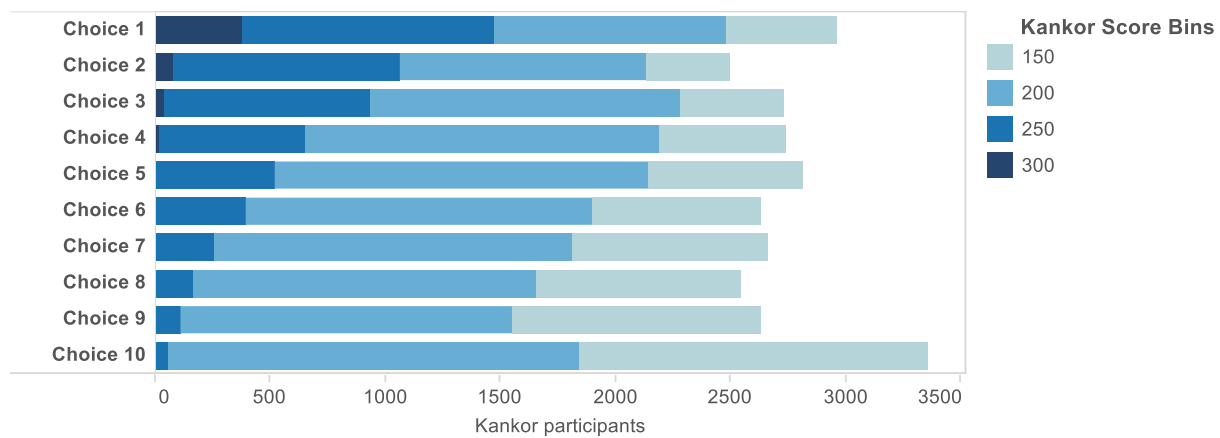

*Figure 4: Assessment of Kankor candidates' performance relevance to the choice of fields study*

Presently, the data from the MoHE does not show the five or ten choices of the candidates i.e. Medicine, Engineering, Computer Science, etc. But if such data is available then questions like what fields of study are popular and trendy among the candidates could be answered. Also, it could be determined if the candidates' choice of fields of study matches systematically and thoroughly. This will give opportunities for policymakers to introduce proper policies and decisions in order to improve the choice of fields of study and the Kankor settings accordingly.

### High School Performance and Grades Assessment

The author made an assessment in order to find out the level of contribution of high school performance and grades in a successful Kankor. The outcome of the analysis depicts that high school scores are not strongly related to success in Kankor since there are public and private high schools whose scoring systems are not trustworthy and realistic (see Figure 5). Therefore, with the current structure, high school performance, and grades are misleading. In addition, from another assessment and evaluation done on the DKRD, it can be concluded that a successful Kankor is not dependent on any particular category of questions i.e. Mathematics, Natural, and Social Sciences, etc. One should get high scores in all the categories in order to get admitted into a favorite university or other higher education institutions.

If the data relevant to the candidate choice of fields of study and score details are stored and produced properly, the author proposes further analyses to be performed to find out other interesting patterns.

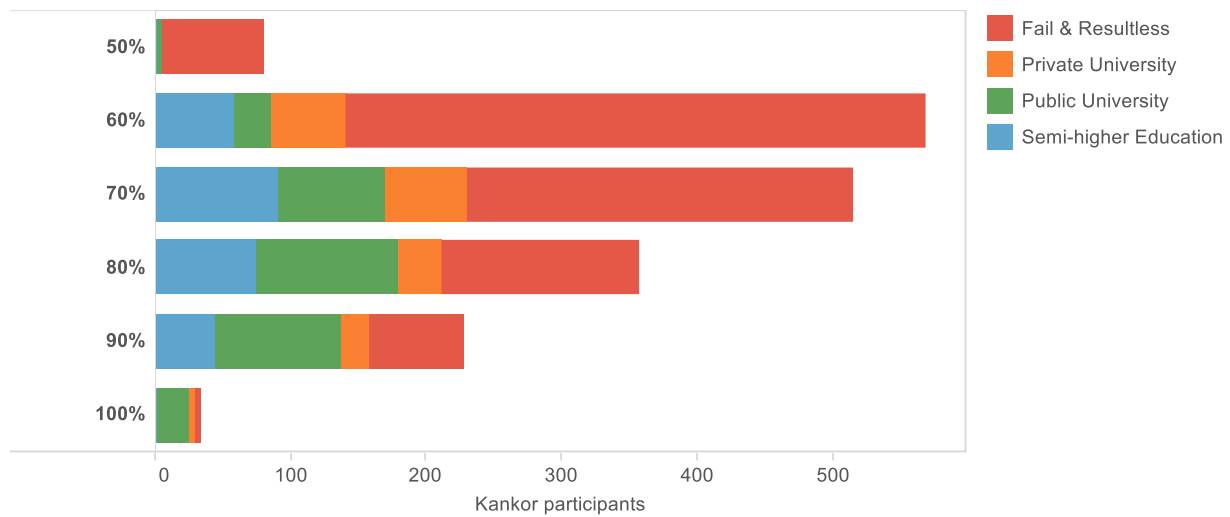

*Figure 5: Contribution of High schools' marks (for grades 10, 11, and 12) in the Kankor*

## Discussions

With the current ineffectual method of recording and producing data, accurate statistics and detailed insights are not produced. The data will neither be more valuable for effective research studies unless they are immensely pre-processed and cleaned. The author proposes the following recommendations for the development of storing and producing effective data as well as improvement of Kankor settings.

1. Attributes with predefined domain values should have proper tags and should be divided into separate attributes. This increases the opportunities to drill down to generate accurate and detailed reports and insights: For example, the attribute "Candidate's Result" should be divided into the following attributes: field of study, educational institution, type of educational institution, location of educational institution, etc.
2. A better approach would be to categorize more or less 150 fields of study (after addressing duplication) into the following main streams: Natural Science, Social Science, Health Science, Literature and Humanities, Islamic Education, Technical and Vocational Education, etc. Also, ensure that the same system is unified across all provinces and the same terminology is used. Currently, in some provinces, the following fields of study: Languages, Mathematics, Physics, Biology, Chemistry, History, Geography, Psychology, Philosophy fall under 'Education'. In some other provinces, they fall under Sciences and Social Sciences. While in some other provinces some of the mentioned fields of study are taught under Humanities and Science faculties. With such a mechanism, unified and transparent reports could not be generated easily.
3. Storing and producing detailed data is substantial: For example, if the candidates' choices of fields of study and their orders are recorded and produced, then the popular fields of study among the candidates could be discovered precisely. Also, assessment could be made if the candidates' choices are systematic i.e. Theology, Fine Arts, Medicine and Engineering are not systematic since they are not relevant to a particular main stream. Additionally, it is recommended to store and produce score details i.e. score for mathematics, natural sciences, social sciences, and languages and also allocation of the number of available seats for higher educational institutions, and other valuable data.
4. The author proposes the following structure with the following attributes: Kankor year, candidate's Kankor ID, candidate's personal and demographic information, gender, cell phone number, candidate's current and permanent province, candidate's high school, high school location (province), high school ownership (public or private), high school geographic (urban or rural), candidate's choice fields of study (1-5), department, field of study, main streams (Natural Science, Social Science, Health Science, Literature and Humanities, and so on), educational institution, educational institution types (public university, private university, teacher training, Islamic education, technical and vocational institutions), educational institution location, educational institutions' available seats, score details, score total, result (Pass, Fail, Fraud, Result-less, and other identifiers), and other supportive and valuable data.
5. The author assumes Kankor Committee of MoHE, through its Kankor Management System, is capable of storing and producing the data as mentioned. If so, unfortunately, the data is not recorded properly and also never shared in detail for research studies. If generations of detailed data costs Kankor Committee, then it is recommended that while they generate Kankor results, they also produce detailed data in excel spreadsheets or some other common format. Then candidate's confidential data could be excluded before sharing with other parties. If the Kankor Management System lacks the feature to store and produce data in detail and in proper method, then, it is highly recommended to keep it in mind for the next version.

## Conclusion

Descriptive analyses reveal that available data are not sufficient to generate accurate and in-depth insights unless they are immensely pre-processed. The author proposed a proper structure to record and produce detailed data of Kankor participants as a solution, which greatly improves the existing structural models.

The current Kankor structure is found not to be the proper medium to identify the level and extent of candidates' competence and skills. For example, according to the author's observation, since Kankor lacks questions to identify Computer and English literacy, the majority of candidates that get admitted into Computer Science do not have enough knowledge of basics of computer and English language which are the cornerstones for the field of Computer Science. Also, analyses illustrate that admission is strongly dependent on the number of available slots at the educational institutions. A candidate can perform well in Kankor and get a high score but still not get into his/her favorite field of study because of the order in which s/he has chosen fields of study. Therefore, the candidate's uninformed choice of field of study might lead to no admission. Furthermore, statistics reveal high schools lack a standard and realistic scoring system; and also the social and political conditions are not favorable. Hence, high school performance and scores do not play a fundamental role in the success of candidates in Kankor.

As a result, direct admission into higher education institutions based on candidates' high school performance and scores is not recommended. On the other hand, elimination of Kankor may be extremely challenging in the short-term. To transition gradually, Specialized Kankor may serve as an intermediate bridge to enhance the existing admission processes. A Specialized Kankor will allow allocation of human capital in appropriate fields of study, leading to a more specialized labor force in the long-term. Results of the questionnaires for university and high school administrators also show that everyone believes that Specialized Kankor is a better means to identify the skills and competence of the candidates. Offering specialized Kankor for every one of the existing fields of study may be difficult, in which case fields of study could be categorized into main streams proposed in this paper.

## Recommendations

Presently Kankor is not able to systematically and methodologically identify skills and abilities of the candidates. The outcome of this research in this context concludes the following recommendations for improvement:

1. It is wise to revise Kankor scoring system based on the candidate's choice of field of study systematically: For example, fields of study relevant to Science require strong understanding and knowledge of mathematics and science subjects. While Theology and similar fields of study do not. This will be a pioneering step leading toward offering specialized Kankor for the main streams.
2. Choosing 5 favorite fields of study out of more or less 150 fields of study is very difficult while the participants do not have sufficient knowledge of them. A better approach would be to categorize them into the following general main streams: Natural Science, Social Science, Health Science, Literature and Humanities, Islamic Education, Technical and Vocational Education, etc.
3. While Kankor results are announced, the MoHE website becomes inaccessible due to traffic loads. The Kankor candidates and their families are impatient to find out about the results. They call friends that have access to the Internet, or they go to internet clubs for hours in order to find out their results. This is a time-consuming process and costs the candidates. With the latest statistics announced, more than 90% of Afghans have access to GSM services via mobile phones. If MoHE introduces a policy to store cell phone numbers of the Kankor candidates which they already do and keep the privacy, then, Kankor result notification could be sent to the candidate's cell phone numbers. This also ensures privacy for Kankor participants instead of sharing Kankor results in excel sheet including all the demographic and other personal information of the candidates on social media, making it a threat against participants.

## Acknowledgements

I am immensely thankful of my supervisors (Prof. Dr.-Ing. Uwe Neumann, Prof. Dr. Sebastian Bab, and Dr. Nazir Peroz) for their constructive comments and feedback. I am also grateful to Bashir Ahmad, a friend of mine who is a graduate of the University of Manitoba, for proofreading this paper. I also appreciate the support of my former graduate computer science students for their support in distribution of the questionnaires in public and private universities and schools of Herat, Afghanistan. I also thank the respondents.

# References

Ali, O. (2015, January 23). Battleground Kankur: Afghan students' difficult way into higher education. Retrieved from https://www.afghanistan-analysts.org/battleground-kankur-afghan-students-difficult-way-into-higher-education/

MoHE, M. of H. E. (2011). *Regulations, rules and procedures of higher education in Afghanistan :: مقرره ها، لوایح و طرزالعمل های تحصیلات عالی افغانستان* (second edition). Kabul: Ministry of Higher Education. Retrieved from http://www.mohe.gov.af/rights/da

Osborne, J. W. (2012). *Best Practices in Data Cleaning: A Complete Guide to Everything You Need to Do Before and After Collecting Your Data*. Thousand Oaks, Calif: SAGE Publications, Inc.

Peck, R., & Devore, J. L. (2011). *Statistics: The Exploration & Analysis of Data* (7 edition). Australia ; United States: Brooks / Cole.

Sakhi, H., & Nabizadah, A. A. (2015). *KANKOR POST-EXAMINATION MIS*. Kabul University, Kabul.

Williams, G. (2011). *Data Mining with Rattle and R: The Art of Excavating Data for Knowledge Discovery* (2011 edition). New York: Springer.